# Disorder effects on exciton-polariton condensates

G. Malpuech, D. Solnyshkov

Institut Pascal, Nanostructures and Nanophotonics group

Clermont Université, Université Blaise Pascal, CNRS, France

**Summary**

The impact of a random disorder potential on the dynamical properties of Bose Einstein condensates is a very wide research field. In microcavities, these studies are even more crucial than in the condensates of cold atoms, since random disorder is naturally present in the semiconductor structures. In this chapter, we consider a stable condensate, defined by a chemical potential, propagating in a random disorder potential, like a liquid flowing through a capillary. We analyze the interplay between the kinetic energy, the localization energy, and the interaction between particles in 1D and 2D polariton condensates. The finite life time of polaritons is taken into account as well. In the first part, we remind the results of [G. Malpuech et al. Phys. Rev. Lett. 98, 206402 (2007).] where we considered the case of a static condensate. In that case, the condensate forms either a glassy insulating phase at low polariton density (strong localization), or a superfluid phase above the percolation threshold. We also show the calculation of the first order spatial coherence of the condensate versus the condensate density. In the second part, we consider the case of a propagating non-interacting condensate which is always localized because of Anderson localization. The localization length is calculated in the Born approximation. The impact of the finite polariton life time is taken into account as well. In the last section we consider the case of a propagating interacting condensate where the three regimes of strong localization, Anderson localization, and superfluid behavior are accessible. The localization length is calculated versus the system parameters. The localization length is strongly modified with respect to the non-interacting case. It is infinite in the superfluid regime whereas it is strongly reduced if the fluid flows with a supersonic velocity (Cerenkov regime).

## I Introduction

In a normal fluid, the viscosity arises because of the elastic scattering of the particles which compose it. This includes both the scattering on the external potential, for example, the walls of the capillary, and the scattering of the particles on each other, if their velocities are different. In contrast to that, for a condensate of weakly interacting bosons (a Bose-Einstein condensate – BEC), which will be the main object studied in this chapter, single independent particles are replaced by collective sonic-like excitations [1, 2]. As a result, such condensate propagating with a velocity smaller than the speed of sound cannot dissipate its kinetic energy by scattering on a disorder potential or on the non-condensed particles. This collective behavior results in a vanishing mechanical viscosity, called superfluidity.

However, sometimes the potential fluctuations can be large enough to destroy the superfluid behavior of a Bose condensate by provoking its complete localization. The question of the interplay between kinetic energy, localization energy, and the interaction between particles has been widely studied in solid state physics since the seminal work of Anderson [3] which described the localization of electrons in a disordered media. Some works have addressed these questions for a gas of bosons in the eighties [4], and this activity took an enormous theoretical and experimental expansion since the observation of the BEC of cold atoms [5]. Particularly interesting to study is the simple case of a 1D weakly interacting Bose gas moving in a disorder potential. Two different model frameworks are typically considered: discrete and continuous. The discrete lattice models usually employ the Bose-Hubbard Hamiltonians by which Mott insulator, Bose Glass, superfluid, or Anderson localized phases are described [6]. The continuous models are usually employed for the description of a relatively weak and smooth potential, where one cannot apply the tight-binding approximation. The theoretical modeling can be performed in this case with the Gross-Pitaevskii equation [7,8 and refs. therein].

Exciton-polaritons are the quasi-particles formed of cavity photons strongly coupled with quantum well excitons, which are expected to behave as weakly interacting bosons, at least at relatively low densities. Despite their short life time, they can thermalize to a quasi-thermal (Bose) distribution [9, 10, 11,12, 13] which can in principle allow the polariton gas to undergo a Berezinskii-Kosterlitz-Thouless phase transition towards a superfluid state [14, 15, 16]. In CdTe or GaN cavities, this superfluid behavior of the condensed phase was not observed because of the presence of a strong in plane disorder which tends to localize the condensate, leading to the formation of a glassy phase [17]. In cleaner GaAs-based samples, the generation of a superfluid is in principle simpler and the observation of a renormalized linear dispersion above the condensation threshold has been reported [18].

Another possible way for generating a polariton superfluid besides the BKT (equilibrium) phase transition is to use the resonant excitation configuration as proposed in 2004 [19] and in 2008 for spinor polaritons [20]. The idea is to pump a polariton state with a laser, which should be slightly blue-detuned from the bare polariton dispersion. If the blue shift induced by the inter-particle interactions in the macro-occupied pumped state exactly compensates the detuning, the laser and the polariton mode become resonant, and the dispersion of elementary excitations is similar to the equilibrium case, and the pump state can propagate as a superfluid. This configuration has been recently used [21, 22] to generate a high density flux of moving polaritons and to study their elastic scattering on a large in-plane defect. A substantial decrease of the flux dissipation by elastic scattering has been observed, but the expected singular character of the superfluid formation under resonant pumping has not been evidenced. However, this type of experiment is really opening a new research field. It reveals the enormous potential of the polariton system to study quantum hydrodynamic effects when a moving quantum fluid hits a large defect (typically larger than the fluid healing length). As predicted [23, 24], this configuration has allowed the observation of oblique solitons [25] (2-dimensional stable solitons [26]), whereas the accounting of the spin degree of freedom allowed to predict the formation of oblique half solitons [27]. Another very promising configuration is given by the fabrication of high quality 1D GaAs microwires [28]. In these samples the radiative particle life time can reach 30 ps, which is one order of magnitude longer than in other material systems. Under non resonant excitation, the 1D character allows the formation of a high density non-equilibrium condensate moving along the wire, spatially independent from the pumping region. Because of the long life time, the propagation for large distances can take place without a substantial decay of the particle density. This is, therefore, a quite ideal configuration, where the motion of a condensate in a random continuous potential can be studied versus the velocity and the density of particles.

In this chapter we do not consider the specific case of a condensate hitting a single potential barrier and we do not study the formation of topological defects, such as solitons. Parametric instabilities are also not taken into account. We consider a stable condensate, defined by a chemical potential, propagating in a random disorder potential, like a liquid flowing through a capillary. This chapter is organized as follows. In the first section, we give an overview, and a critical discussion of the literature devoted to the disorder effects on polaritons. In the second section, we recall the main expected properties of a static Bose Einstein condensate placed in a disorder potential, analyzing the interplay between localization and interaction effects. In the last section, we finally consider a propagating condensate, first in the linear non-interacting limit, mainly discussing Anderson localization of polaritons, and then taking into account the interactions. We then discuss the critical condition required for the occurrence of superfluidity. We finally address the question of the interplay between the kinetic energy, the localization energy, the interaction between particles, and the lifetime. To summarize the different possibilities for our interesting system, we plot a phase diagram.

## II Historical overview

The role of the structural disorder on the linear optical properties of microcavities was first evidenced by Resonant Rayleigh Scattering experiments. An exciton-polariton eigenstate is normally characterized by a wave vector $\vec{k}$ which defines the polariton energy through the dispersion relation $E(k)$. Because the polariton is a mixed state of two particles having different masses, its dispersion is not parabolic. If the in-plane translational invariance is broken by the presence of disorder (acting on one or both polariton components), the polariton wave $\vec{k}$ is not anymore a good eigenstate. Such wave, for example resonantly created by a laser, scatters toward the "elastic ring" of isoenergetic polariton states as shown experimentally in [29]. This perturbative description is correct if the disorder amplitude is small with respect to the kinetic energy. If the kinetic energy is small compared with the disorder amplitude, the particles become strongly localized which provokes a strong change of the particle dispersion, as discussed for example by Agranovich and Litinskaia in 2001 [30]. In the hypothetic case where the decay processes such as life time, or phonon scattering are negligible, this process leads to an iso-distribution of particles on the elastic ring. If the coherence is sufficient, this process should lead to a weak localization of the polariton waves called Anderson localization. So far this process has not been yet neither observed, nor described in polaritonic systems. It is typically dominated by the short radiative life time of particles which limits the spatial extension of a polariton wave much more than localization effects. The other consequence of the effect of disorder is the inhomogeneous broadening of the polariton line. As a result, the sum of the widths of the lower and upper polariton modes is not constant versus the exciton-photon detuning, but shows a minimum [31]. This result was interpreted in the 90's as a "motional narrowing" effect which led to some controversy [32]. Another important aspect relies on the type of material used to grow the structure which strongly affects the amplitude of the disorder potential. GaAs based samples show the best structural qualities with the inhomogeneous broadening of the polariton line which can be as low as 0.1 meV. InGaAs QWs are a bit less good, with values of the order of 0.5 meV. In CdTe-based structures the inhomogeneous broadening value is typically a few meVs. It is typically 10 times larger in GaN based samples, and again about 5-10 times larger in organic based structures. Disorder affects both the excitonic and photonic parts of the polariton modes, but the typical correlation lengths for both are different.

After the study of linear properties of the microcavities, the non-linear optical response of microcavities has been explored under resonant and non-resonant excitation. Under non-resonant excitation, the goal of experimentalists was to achieve polariton lasing, first suggested by Imamoglu in 1996 [33]. A non-resonant laser pulse creates high energy electron-hole pairs which bind to form an incoherent exciton reservoir which in turn forms an exciton-polariton condensate in the ground state. The polariton condensation is possible because of exciton-exciton and exciton-phonon interaction. Because of the finite polariton life time and limited efficiency of the relaxation processes, the polariton condensation is in principle an out-of-equilibrium process. However, different regimes can be distinguished depending on the type of materials used, on the exciton-photon detuning, polariton life time and on the size of the pumping spot. A thermodynamic regime can be defined [11], corresponding to the achievement of a quasi-thermal distribution function. In such a case, important features, such as the critical condensation density, or the polarization of the polariton condensate can be extracted from thermodynamic calculations, which often have the advantage of being analytical. On the other hand, another regime, called kinetic, does exist as well, where the condensate features are fully governed by the dynamics of the system. This feature and the existence of the two regimes in a given structure with a possible transition between them have been demonstrated in all types of semiconductor microcavities: CdTe [11], GaAs [34], GaN [13]. Technically, the first clear evidence of the feasibility of the Imamoglu's proposal has been published by Le Si Dang in 1998 [35]. However, the work, which is now mostly cited and recognized as being the one where polariton condensation was observed, is the Nature paper of 2006 by Kasprzack et al. [11]. The sample and the experiment performed were the very same as in 1998, but three new measurements were added. First, the distribution function was measured and found to be close to an equilibrium distribution function. Second, the spatial coherence was found to pass from 0 to 30 % at distances of about 5-10 μm. Third, the condensate was found to be linearly polarized above threshold, which is another confirmation of the condensation taking place, because the polarisation is the order parameter of such phase transition in a spinor system [36]. With these new data, the observation of polariton condensation close to thermal equilibrium ("polariton BEC") was claimed. Since that time, there is a strong tendency to state that polariton condensation is a non-equilibrium process and that the achievement of equilibrium (which was one of the important results of [11]) is unimportant or impossible. If this is indeed true, it would be probably fair to slightly rewrite the history. The build up of linear polarization, pinned along crystallographic axis in the polariton laser regime was demonstrated before the Ref. [11], by the group of Luis Vina [37]. Also, a "non-equilibrium" condensation had been reported earlier, in two papers of 2005 by M. Richard et al. In Ref. [38], condensation took place in finite-k states because of the use of a small pumping spot, as it was understood later. The coherence between different k-states was evidenced directly. In Ref. [39], condensation was taking place in the ground state, but was stated as non-equilibrium, because of the use of a pulsed pumping laser. The condensate was found to be spatially localized in several different spots linked with the presence of an in-plane disorder potential. However, the angular width of the emission remained narrow, well below the diffraction limit imposed by a single emitting spot. This evidenced that these different spots were emitting in phase together, demonstrating the onset of spatial coherence, one year before it was made by interferometric technique in the Nature of 2006. The most convincing evidence of the build up of a spatially coherent condensate was given later, in the Nature Physics of 2010 by Wertz et al [28], where a coherence degree larger than 80 % was found for distances over 200 micrometers (50 times the De Broglie Wavelength). In this last work however and similarly in the PRL of 2005 [38], the condensates generated are completely out of equilibrium because of the use of a small pumping spot which limits the overlap between the thermal exciton reservoir and the polariton condensate itself.

The conclusion one can draw from this brief historical overview is that if the achievement of quasi-equilibrium is absolutely uninteresting, as suggested in many recent works, then possibly other works than Ref. [11] could be cited as the first evidence of the polariton condensation achievement, depending on the importance given to the achievement of spatial coherence. One could interestingly notice that spontaneous symmetry breaking is often referred to as the "smoking gun" of Bose condensation. From that point of view, the clearest evidence of polariton condensation could be the J.J. Baumberg's PRL of 2008 [40], where the build-up of a condensate polarisation above threshold was observed, with a polarization direction varying randomly from one experiment to another, and not pinned to any crystallographic axis.

Let us now go back to the main topic of this chapter, which is the effect of disorder on polariton condensates. From this point of view, the experiments performed on CdTe-based samples are really of strong interest, because of the relatively large disorder present in these samples. In [39] already, the condensate was found to be strongly inhomogeneous in real space, peaked around the in-plane fluctuations of the potential. The formation of vortices pinned to these defects was already suggested, which, we remind, was later demonstrated clearly in the Science [41]. In the Nature paper [11], evidences of localization were even stronger, with the appearance of flat dispersion around $k$=0.

**III Static condensate in a disorder potential**

Let us first consider a static gas of bosons at 0 K in a random potential as discussed in [17]. A realization of this potential is shown on the figure 1. The ground state of this system can be found by solving the Gross-Pitaevskii equation by minimization of the free energy for a fixed number of particles. In a non-interacting case, at thermal equilibrium all particles are strongly localized in the deepest site, so-called Lifshits state. In practice, thermal activation or the out-of-equilibrium system character allow the occupation of many localized states having possibly different energies (Lifshits tail). Such case is shown on the figure 2a and 2b demonstrating the spatial distribution of particles and their dispersion, which is quasi-parabolic near the ground state, characterized by substantial inhomogeneous broadening. If one considers a weakly interacting Bose Gas, it fills all localized states having their energy below the chemical potential values μ. The condensate forms independent lakes separated by potential barriers (figure 2c) and forms a glassy phase.

This glassy phase of a Bose condensate can be called a Bose glass, but this term is usually applied to the glassy phase formed due to excessive inter-particle interactions, which inhibit the hopping between different sites in lattice models, which is never the case for polaritons for reasonable parameters. Our glassy phase, therefore, has to be called Anderson glass, and the distinction between the two glasses has become clear since Ref. [6], where the transitions between all four possible phases (Anderson glass, Superfluid, Bose glass, Mott Insulator) were studied.

Even if the system is not in thermodynamic equilibrium and several independent condensates are formed in the potential minima, depending on the particle density the neighboring sites having non-connected non-interacting ground states can start to overlap to finally synchronize, as shown in [42]. However, in 2D the randomness of the potential forbids the creation of a conduction band for the particles until the classical percolation threshold is reached. The spectrum of elementary excitations of the condensate is shown on the figure 2d. It shows the flat dispersion observed experimentally. The superfluid fraction, calculated using the twisted boundary condition technique, is close to zero. This picture corresponds to the one experimentally observed. If one increases the density further, the classical percolation threshold is reached and a fraction of the fluid becomes superfluid as demonstrated by the linear dispersion of the elementary excitations shown on the figure 2f.

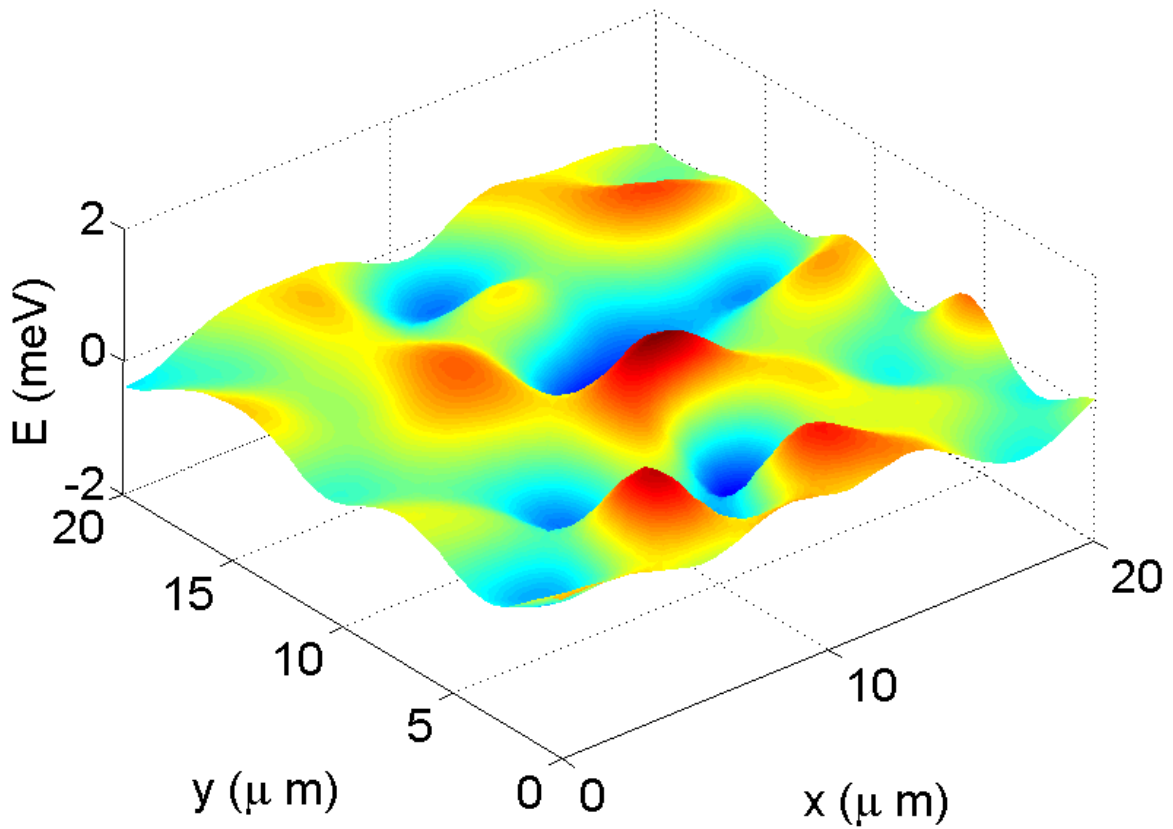

Figure 1. A typical disorder potential characterized by rms fluctuation V0=0.5 meV and correlation length lc=2 μm.

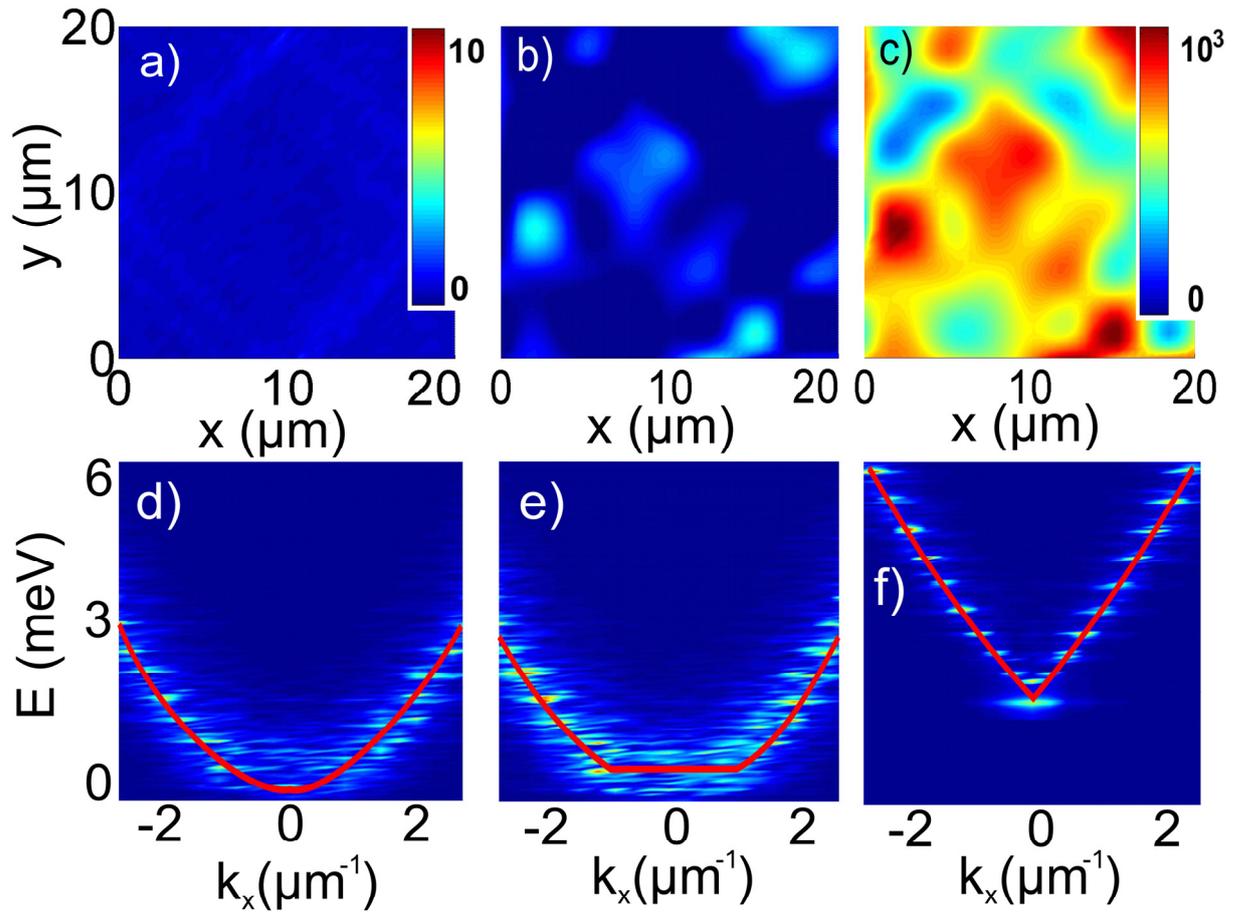

Fig. 2. Spatial images (top panel) and quasiparticle spectra (bottom panel) for a realistic disorder potential. The figures shown correspond to densities 0, 6x10^10 and 2x10^12 cm. The red lines are only guides to the eye, showing parabolic, flat and linear-type dispersions. Colormap of the panel (b) is the same as (c).

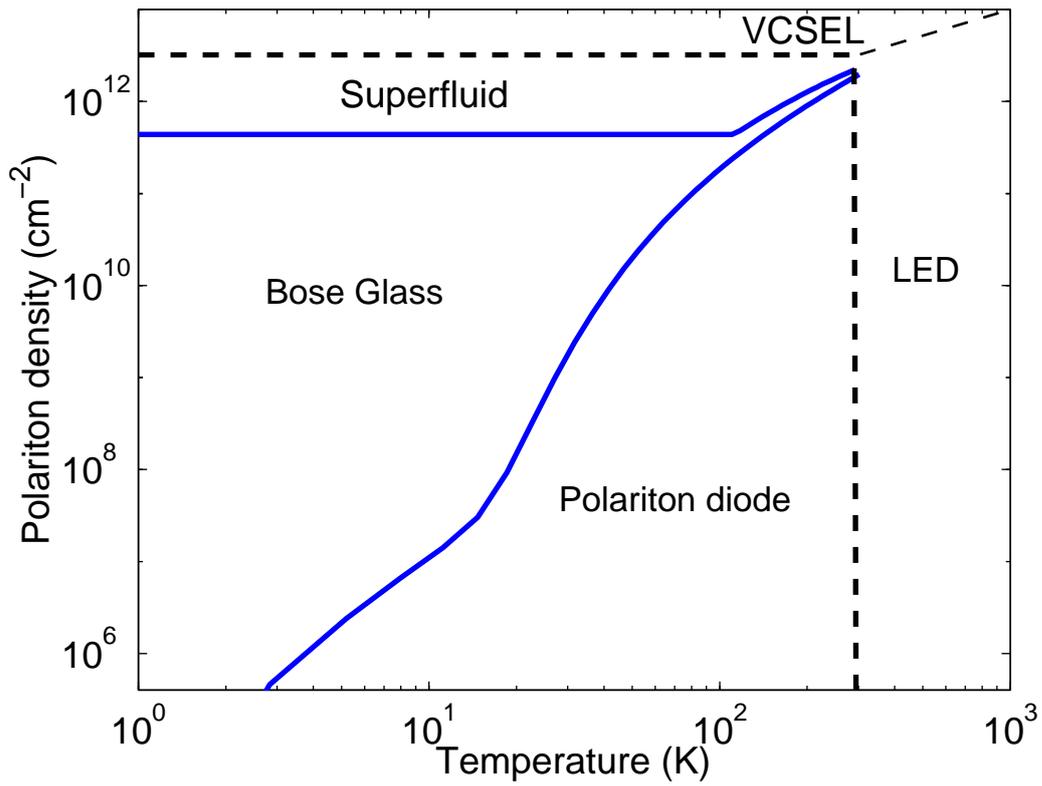

Fig. 3. Phase diagram for exciton-polaritons in a disordered microcavity in the thermodynamic limit. The Bose glass phase can also be called Anderson glass. Superfluidity is only possible above certain density, depending on the disorder rms fluctuations.

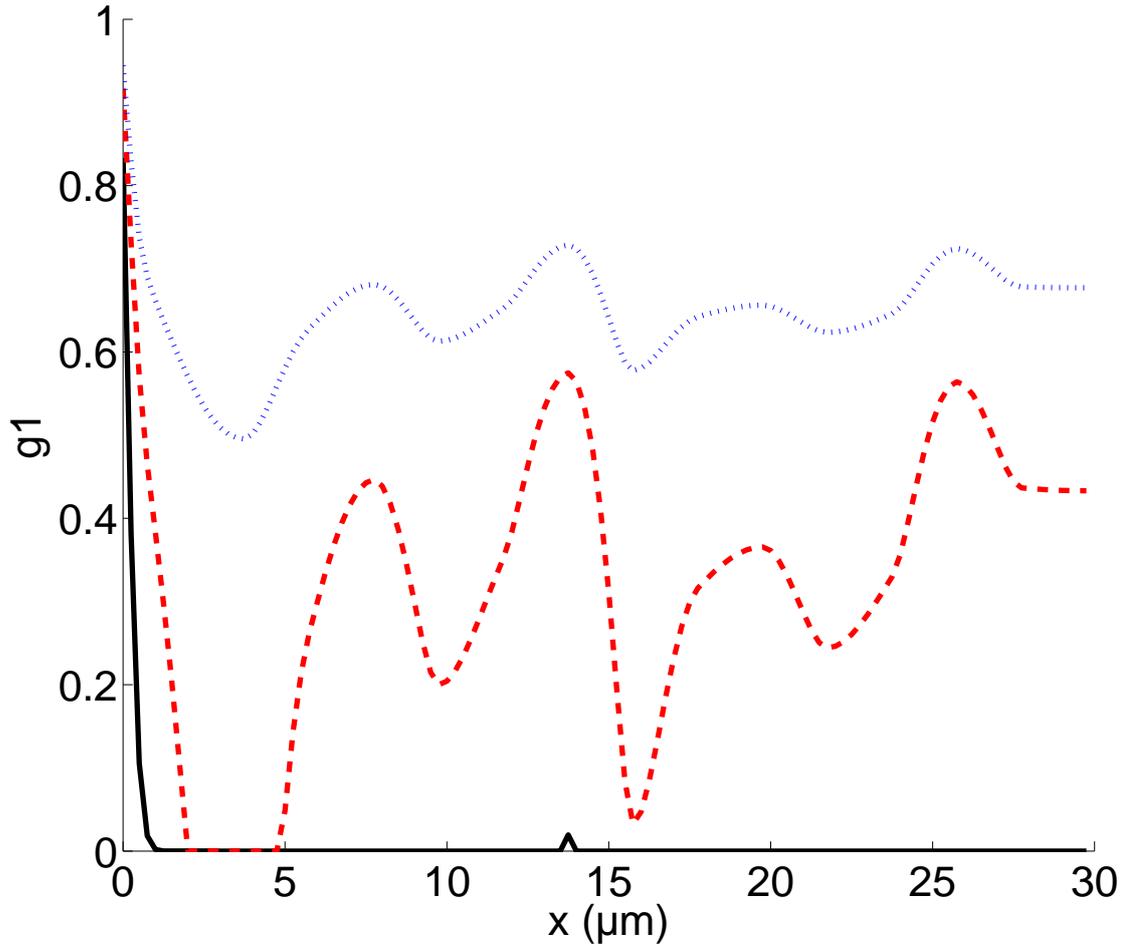

Figure 4. Coherence degree for 3 different condensate densities. In the linear case (almost zero density) the coherence length is given approximately by the de Broglie wavelength. The condensed part is assumed to be 100% coherent, but its density varies from point to point due to the presence of the disorder potential.

A phase diagram from Ref. [17] summarizing these results is shown in figure 3. It demonstrates that exciton-polaritons in a disordered microcavity can present several different phases, and the superfluidity can be recovered only above certain density threshold, linked with the rms amplitude of the disorder. Since the polariton density is limited by the possible loss of the strong coupling, only relatively good material systems with low disorder, such as GaAs, can be expected to show signs of superfluidity.

In the figure 4, we show a calculation of the first order coherence for a finite size 1D system, assuming a thermal uncondensed fraction of particles with a homogeneous spatial distribution (high kinetic energy of uncondensed particles provides their weak sensitivity to the disorder). The coherence degree fluctuates, showing maximum values at the potential minima, because the density of the condensate is higher in these minima. This is again in good agreement with the last work of the Deveaud's group on the very same CdTe sample [43]. The density and coherence fluctuations of the condensate associated with the presence of the disorder potential is measured, and related to the formation of a Bose glass phase (which, we remind, should rather be called

Anderson glass). For some puzzling reasons, the authors insist that this effect is the result of the strong non-equilibrium nature of their system whereas it is obviously not the case.

Similar works, considering 1D condensates in disordered systems have then been published by the group of Vincenzo Savona [7,8]. They reached the same conclusion that the insulating to superfluid phase transition occurs at the percolation threshold. However, one should notice that in an infinite 1D disorder potential with a Gaussian distribution, there is no percolation threshold. This transition therefore only exists either in a finite system, or in a system with a well defined higher bound for the potential fluctuations.

One should point out that, in 2007, an alternative explanation to the absence of superfluidity in polariton condensates was proposed both by the Cambridge group of P. Littlewood [44] and the Trento group of I. Carusotto [45] who attributed the formation of a diffusive (flat) dispersion to the non-equilibrium nature of the polariton codensate. It is indeed well known that in the *resonant* pumping case, the external laser driving the system strongly affects the dispersion of the excitations of the driven mode which can be diffusive, parabolic, or even linear for some precise value of the pumping intensity [19, 20]. In [45], a simple reservoir-condensate model was proposed which gave essentially the same result as the resonant pumping scheme, being very similar to the latter from the mathematical point of view. Depending on the decay time constant of the reservoir, the dispersion of the elementary excitations of the condensate can be parabolic, linear, and, when using an unphysically short decay time for the reservoir (5 times shorter than for the condensate), diffusive. This result is mathematically interesting, but it is clearly completely unrelated to the observed experimental result. The latter statement is also supported by several experimental studies. In [18], a GaAs structure less affected by disorder than CdTe sample has been studied. The dispersion of elementary excitations was claimed to pass from parabolic to linear above threshold. The data could be judged not so convincing, but it is however clear that the dispersion was not diffusive in that case. Finally, tons of experimental works demonstrating the strong disorder effect in CdTe cavities came. One can mention the observation of the condensation in a single potential trap of a few microns by D. Sanvitto et al. [46], or the observation of a Josephson-like dynamics between two localized states where the condensation is taking place [47]. Also, observation of vortices [48] and half vortices [41] was made possible because of the pinning of the flux by the disorder potential.

### IV Localization and superfluidity of a moving condensate

In this second section we discuss a bit more subtle problem of a moving fluid, interacting or not, flowing against a defect or through a disorder potential. This has been a very active field of research since the seminal work of Anderson [3]. It is by far beyond our capacities to give a complete review on this topic and we are going to concentrate on a few specific results.

#### a) Anderson localization

Brownian motion of classical particles in a random potential leads to diffusive dynamics. A particle freely propagates for a typical distance (mean free path, a distance between two impurities) and is scattered in a random direction. The width of the particle distribution around the initial position increases as $\sqrt{Dt}$, where $D$ is the diffusion coefficient and $t$ the time. The probability distribution therefore infinitely broadens in time. In his initial work, Anderson considered a 3D lattice model. The energies of the sites are random. They are connected by some matrix element of hopping between the sites. Anderson shown that in such a model, the wave function of a single quantum particle does not always spread with time but sometimes remains exponentially localized on the initial site. The transparent explanation given by Anderson is that the electron cannot hop to the neighboring site if the energy mismatch exceeds the hopping matrix element. In other words, the diffusion coefficient is zero. In 3D, the existence of a mobility edge has been demonstrated. However, in 1D and 2D any weak random potential is expected to localize a quantum particle, whatever its kinetic energy, unless some extra terms are added to the Hamiltonian, such as the spin-orbit coupling or the interactions. Localization has been evidenced for light waves[49], microwaves [50], sound waves [51], and matter waves [52]. Anderson localization is strongly associated with the concept of phase coherence as we are going to explain below, which is not evident from the initial interpretation given by Anderson.

Let us consider a particle moving in a random 1D potential. This particle hits a defect and is either transmitted or reflected with certain probabilities. In a classical scheme, the reflected fraction will follow a classical random walk, but at a later time it is expected to collide with the same barrier again and participate in the transmission. For a quantum particle, this random walk will lead to destructive interferences, and the reflected part does not give any contribution to the transmission. The wave is therefore attenuated at each barrier, which leads to an exponential decay of the wave function. The effect is really subtle, because the role of interference is to cancel all multi-reflection channels for the transmission, which play a crucial role in the classical diffusive motion.

If one consider 1D systems, a standard approach used [53,54] to calculate the localization length is the so-called "phase formalism" [55], where the Schrödinger equation is solved in a Born approximation using a localized wave function Ansatz. Here, for the sake of demonstration, we choose to use a slightly different approach.

We consider a propagating plane wave with a wave vector $k_0$. The bare dispersion of the particles $E(k_0)$ is assumed to be quasi-parabolic. The wave is propagating in a continuous disorder potential $V(\vec{r})$ with a Gaussian random distribution of energies with a root mean square fluctuation $V_0$ and a correlation length $L_c$. The action of this potential on the propagating particles is treated as a perturbation. It does not provoke any change in the dispersion of the particles, but simply induces some elastic (Rayleigh) scattering processes which are assumed to provoke a decay of the number of the propagating particles (scattering to the other $k$ states). This assumption is based on the fact that, in the Anderson picture, the backscattered part of the wave will never interfere again constructively with the propagating wave.

From the Fermi golden rule, the decay rate of the propagating particles reads

$$\frac{1}{\tau} = \frac{2\pi}{\hbar} \sum_f |M|^2 \delta(E_i - E_f) \qquad (1)$$

where $M = \langle \psi_k | V_{k-k'} | \psi_{k'} \rangle$, $|k| = |k'|$, and $V_k$ is the Fourier transform of the potential. The final result essentially depends on the density of final states, which depends itself on the dimensionality of the problem:

$$\frac{1}{\tau_d} = |x|^2 \frac{2^{d/2} L_c^d}{\hbar} V_0^2 D_d \left( E = E(k_0) \right) \qquad (2)$$

where $d$ is the dimensionality, $k_0$ is the wave vector, $|x|^2$ the exciton fraction. $D^d \left( E = E(k_0) \right)$ is the density of states at the energy of the propagating wave.

From the equation (2) one can extract the inverse of the propagation length as

$$\gamma_d (k_0) = \frac{1}{v \tau_d} = \frac{m}{\hbar k_0 \tau_d} = |x|^2 \frac{2^{d/2} m L_c^d}{\hbar^2 k_0} V_0^2 D_d \left( E = E(k_0) \right) \qquad (3)$$

In 1D and 2D, and for parabolic dispersions, this formula reads:

$$\gamma_{1D} = |x|^2 \frac{\sqrt{2} m^2 L_c}{\hbar^4 k_0^2} V_0^2 \qquad (4)$$

$$\gamma_{2D} = |x|^2 \frac{2 m^2 L_c^2}{\hbar^4 k_0} V_0^2 \qquad (5)$$

The formula (4) is similar to formula (13) of [53], derived from the phase formalism, except that our numerical prefactor $\sqrt{2}$ is replacing $\frac{\pi}{2}$.

In a system with a finite lifetime, a supplementary decay source should be considered and the total decay rate can be written as:

$$\gamma_d (k_0) = |x|^2 \frac{2^{d/2} m L_c^d}{\hbar^2 k_0} V_0^2 D_d \left( E = E(k_0) \right) + \frac{m \left( 1 - |x|^2 \right)}{\hbar k_0 \tau_R} \qquad (6)$$

where $\tau_R$ is the photon life time in the cavity. The scattering of excitons by phonons is neglected, but it could be easily introduced, that would bring additional temperature dependence into the formula. The figure 5 shows, in blue and red for 1D and 2D respectively, the localization lengths of polaritons for a rms potential fluctuation of 0.25 meV and a correlation length of 1 micron. The Rabi splitting is taken 20 meV and the exciton-photon detuning is zero. The cavity photon life time is taken 30 ps. The black line shows the localization length induced by the life time. For these parameters, Anderson localization of photons can be evidenced both in 2D and 1D, despite the considerable effect of the life time. The 1D and 2D cases do not appear extremely different. Particles are more localized in 1D than in 2D for small energies. The situation is opposite for higher energies. The exact position of the crossing of the two curves depends on the coherence length of the disorder.

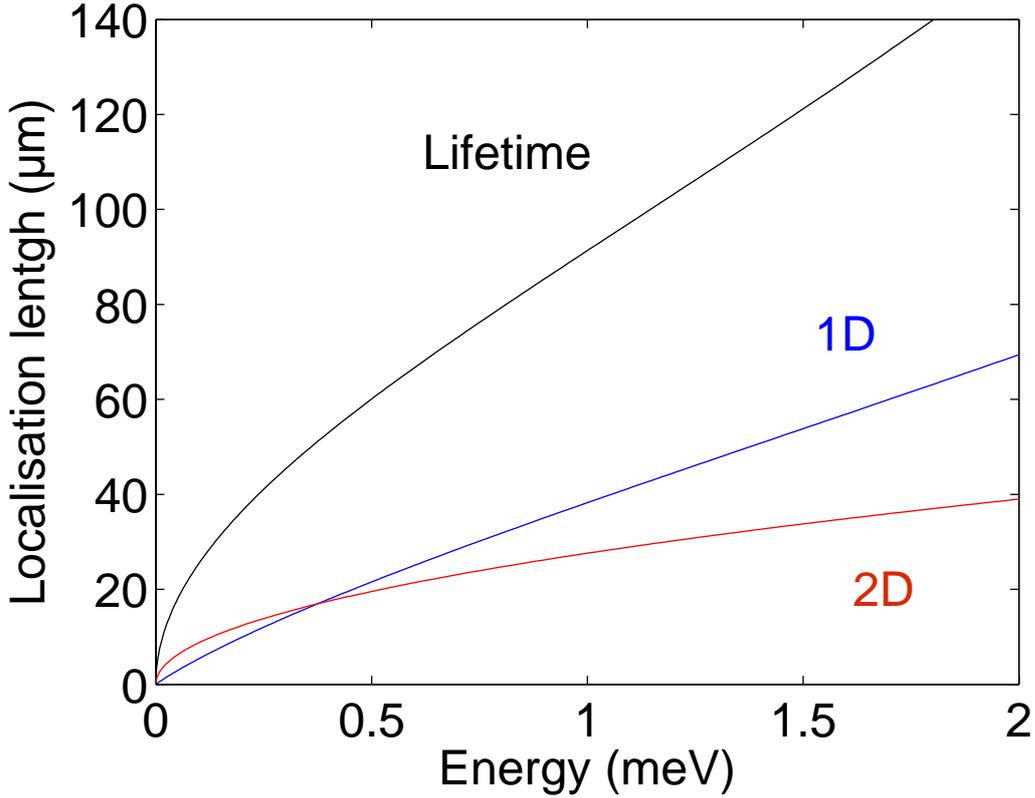

Figure 5: Localization length $\gamma_d^{-1}$ of the propagating polariton condensate (from formula 6). The black curve shows the localization length in a disorder-free structure (life time only).

**b) Superfluidity**

A superfluid is a fluid moving without mechanical viscosity. As first understood by Landau [1], this extraordinary property can arise in a fluid if the dispersion of the elementary excitations of the fluid is linear. In that case, the fluid can flow without friction as long as its velocity remains below the speed of sound $c_s$ given by the slope of this linear dispersion. In 1947, Bogoliubov [2] established the link between superfluidity and condensation by showing that the dispersion of the excitations of weakly interacting bosons follows, at low $k$, the linear shape proposed by Landau. Physically, the suppression of viscosity is associated with the suppression of elastic scattering processes provoked by the roughness of the surface along which the fluid in propagating.

Below we will demonstrate the Bogoliubov linearization procedure for a condensate of weakly interacting bosons propagating with a certain wave vector. Let us consider the Gross-Pitaevskii equation :

$$i\hbar \frac{\partial \psi(r,t)}{\partial t} = T\psi(r,t) + \alpha |\psi(r,t)|^2 \psi(r,t) \qquad (7)$$

where $T$ is the kinetic energy operator, namely $-(\hbar^2/2m)\Delta$ for a parabolic dispersion. A plane wave $\psi_0 e^{i(k_0 r - \mu t)}$ is a solution of the equation (7) with :

$$\mu = \omega_0(k_0) + \alpha |\psi_0|^2 \tag{8}$$

where $\omega_0(k_0)$ is the kinetic energy.

We then consider a small perturbation of the condensate wave function:
$$\psi(r,t) = \psi_0 + \delta\psi.$$

The non linear term of (1) reads:
$$|\psi(r,t)|^2 \psi(r,t) \approx |\psi_0|^2 \psi_0 + 2|\psi_0|^2 \delta\psi + 2\psi_0^2 \delta\psi^*$$

which means that the non-linear term couples $\delta\psi$ and $\delta\psi^*$.

Therefore, the solutions of the equation 7 can be written as

$$\psi = \psi_0 e^{i(k_0 r - \mu t)} \left( 1 + A e^{i(kr - \omega t)} + B^* e^{-i((2k_0 - k)r - \omega t)} \right) \tag{9}$$

Inserting (9) into (7), neglecting high order terms yields the linearized system of equations below :

$$A(\mu + \omega) = \left( \omega_0(k) + 2\alpha |\psi_0|^2 \right) A + \alpha \psi_0^2 B$$
$$B(\mu - \omega) = \left( \omega_0(2k_0 - k) + 2\alpha |\psi_0|^2 \right) B + \alpha \psi_0^{*2} A \tag{10}$$

The solutions of the system (10) are given by:

$$\omega_\pm = \frac{-\Delta_1 - \Delta_2}{2} \pm \sqrt{\left( \frac{\Delta_1 - \Delta_2}{2} \right)^2 + \alpha |\psi_0|^2 (\Delta_1 - \Delta_2)} \tag{11}$$

where
$$\Delta_1 = \omega_0(2k_0 - k) - \omega_0(k_0)$$
$$\Delta_2 = \omega_0(k_0) - \omega_0(k)$$

For a parabolic dispersion for $k \sim k_0$

$$\omega_\pm = \left( \frac{\hbar k_0}{m} \pm \sqrt{\frac{\alpha |\psi_0|^2}{m}} \right) \hbar(k - k_0) \tag{12}$$

$$|A_\pm|^2 \equiv \frac{\left( \omega_\pm - \omega_0(k_0) + \omega_0(2k_0 - k) + \alpha |\psi_0|^2 \right)^2}{\left( \left( \omega_\pm - \omega_0(k_0) + \omega_0(2k_0 - k) + \alpha |\psi_0|^2 \right)^2 + \left( \alpha |\psi_0|^2 \right)^2 \right)}$$

$$|B_\pm|^2 \equiv \frac{\left(\omega_\pm + \omega_0(k_0) - \omega_0(k) - \alpha|\psi_0|^2\right)^2}{\left(\left(\omega_\pm + \omega_0(k_0) - \omega_0(k) - \alpha|\psi_0|^2\right)^2 + \left(\alpha|\psi_0|^2\right)^2\right)} \qquad (13)$$

The figure 5 shows the dispersions of elementary excitations of a polariton condensate propagating with a wavevector $10^6$ m$^{-1}$. Panel (a) illustrates the Čerenkov regime, where the elastic scattering towards an excited state with the same energy as the condensate is possible because the renormalization of the dispersion due to the interactions is not strong enough. Panel (b) corresponds to the superfluid case, where the dispersion is so strongly renormalized that the elastic scattering towards any excited state is no more possible, and therefore, due to the Landau criterion, the viscosity of the propagating polariton fluid drops to zero.

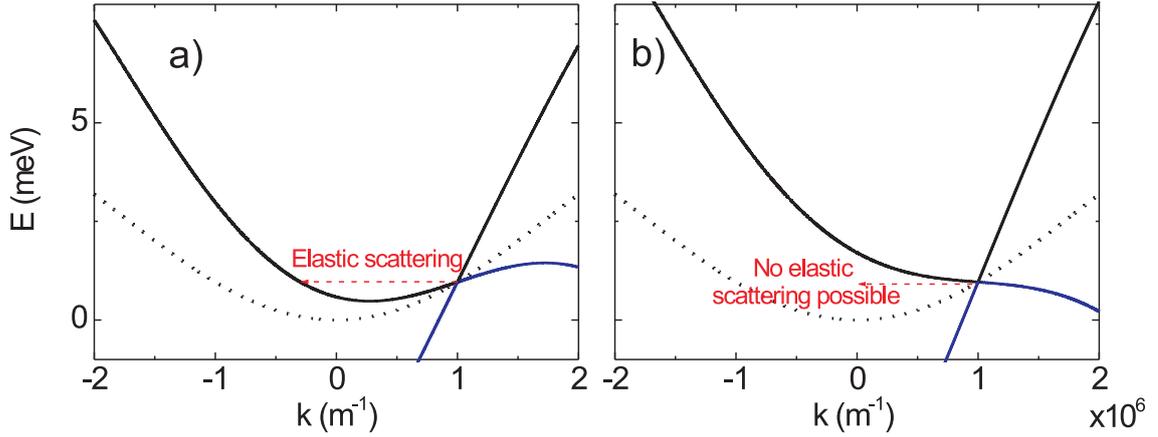

Figure 6. Bare polariton dispersion (black dotted line) and the dispersions of elementary excitations (black and blue/light grey solid lines) in the Čerenkov (a) and superfluid (b) regimes. Elastic scattering towards excited states is indicated by a red arrow.

The solution (12) tells us that the excitations of the condensate are created by a kind of parametric process symmetric with respect to the energy and the wave vector of the condensate. However, when the final state is far from the initial dispersion (far means that the energy difference becomes comparable or larger than $\mu$), the amplitude of the corresponding state vanishes, which means that this state does not exist and that the excitation is just a plane wave (free particle), essentially unperturbed by the presence of the condensate. The condensate itself is expected to be stable against parametric processes if $\omega_\pm$ are real values, which is always the case for a parabolic dispersion in 1D.

OPO-like processes are very well known in the polariton community because of the non-parabolicity of the polariton branch which makes them resonant. They have therefore been strongly studied since the year 2000 and the pioneering works of Savvidis [56], Stevenson [57], and the strong theoretical contribution of Ciuti [58, 59]. After the claiming by the Yamamoto's group of the observation of this linear Bogolubov dispersion [18], some doubts came from the fact that they did not observe the negative part of the dispersion. Recently the Lausanne group attacked this question, by trying to observe this "ghost branch" by heterodyne detection [60]. One should notice that off resonant emission coming from "ghost branches" has already been evidenced [59] under resonant excitation. In fact, even if these new branches (eigen values) come from OPO-like processes, such processes are not balanced. The figure 7 below shows the expected emission spectra for the same parameters as for figure 6. The dispersion is indeed perturbed, but the "ghost branch" is not easily visible.

The second comment is that superfluidity is fully linked with inter-particle interactions. A non-interacting gas moving with any velocity in a weak random potential gets exponentially localized because of Anderson localization. If interactions are considered, the same gas can show a superfluid behavior which means that it does not even "feel" the presence of the disorder during the propagation.

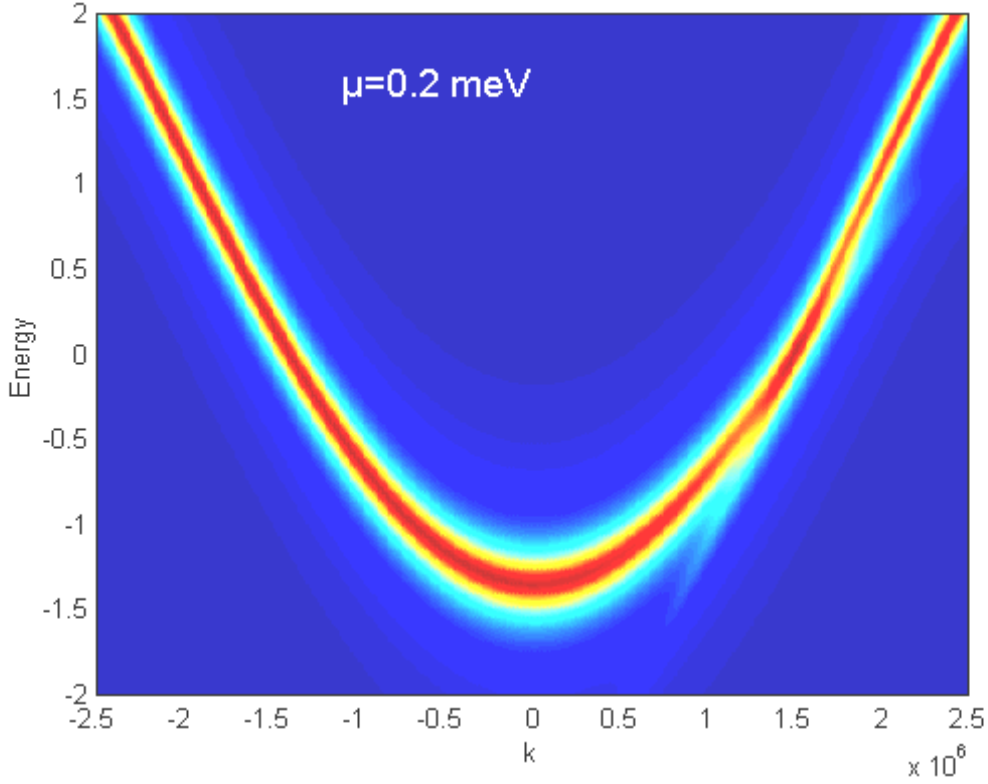

Figure 7: Emission from the elementary excitations of a polariton condensate, calculated using formula 11 and 13. The condensate has an in-plane wave vector $1.5 \; 10^6$ m$^{-1}$. The interaction energy $\alpha |\psi_0|^2$ is 0.2 meV. All states are assumed to emit light proportionally to the coefficients calculated with the formula 13, including a linewidth of 0.1 meV. The "Ghost Branch" can be guessed to the right of the main dispersion.

**c) Phase diagram**

In what follows, we consider a 1D condensate moving in a random disorder potential. This condensate is characterized by a fixed energy, which means that we neglect the possible occurrence of parametric instabilities, and the possible formation of topological defects, such as solitons. This is a strong approximation. In [61] for example, such situation is considered numerically and the condensate is found "unstable" in a wide range of parameters, which means that the propagating flux cannot be characterized by a single frequency. Therefore, in what follows, we assume that the only possible dissipative process for the propagating condensate are backscattering processes, which could also be called "emission of Čerenkov waves". The qualitative picture we propose is summarized on the phase diagram shown on the figure 3. As said in the previous section, at negligible interaction energy, bosons in a 1 D system are known to be always localized in a random potential, whatever their kinetic energy. Two different localization regimes can however be distinguished. If the kinetic energy $\omega_0$ is much smaller than the amplitude of the random potential $V_0$, the condensate is classically localized in some independent localised states, referred to as Liftshits states in [53]. The localisation length is typically given by the correlation length $L_c$ of the disorder potential. If $\omega_0 > V_0$, the bosons are still localised but the physical mechanism is more subtle, relying on Anderson localization, as discussed in the previous sections. A propagating plane wave is backscattered by the disorder with some finite scattering time proportional to the density of states at the propagating energy $\tau_{1D} = \frac{\hbar^3 k_0}{\sqrt{2} m L_c V_0^2}$, where $k_0$ is the wave vector of the wave, $m$ - the mass and $L_c$ - the correlation length of the disorder potential. The localization length is given by $\frac{\hbar k_0}{m} \tau_{1D}$ and is no more directly proportional to the coherence length of the disorder potential. The smooth crossover between the two regimes takes place approximately when $\omega_0 = V_0$.

This picture is dramatically modified by the interactions. Their first role is to bleach the disorder potential, whose effective height is given by $V_0 - \alpha |\psi_0|^2$. This reduces the kinetic energy at which the transition between strong and Anderson localization takes place to $\omega_0 = V_0 - \alpha |\psi_0|^2$. Second, the interactions provide a finite superfluid velocity $v_s = \sqrt{\frac{\alpha |\psi_0|^2}{m}}$. A condensate propagating with a velocity smaller than $v_s$ shows no viscosity because of the suppression of the elastic backscattering – the mechanism, which is also responsible for the occurrence of the Anderson localization. However, if the system is strongly localized, it cannot be called superfluid; because there is no motion of the fluid in this case. Therefore, superfluidity can take place only if the kinetic energy of the condensate is large enough for the particles not to be strongly localized, but small enough to have the condensate velocity below the speed of sound in the media, which is expressed by the condition $V_0 - \alpha |\psi_0|^2 < \omega_0 < \alpha |\psi_0|^2 / 2$. If $\omega_0 > \alpha |\psi_0|^2 / 2$, a superfluid enters the so-called Čerenkov regime. We show that in this regime, the backscattered amplitude is enhanced with respect to the non-interacting case, providing a reduced localization length.

An important remark regards the difference of this model with the one describing disordered lattices. In the lattices, the coupling between regularly spaced localized states can result in the formation of a conduction band allowing superfluidity. In the case of a continuous random potential, the delocalization threshold corresponds to the percolation of the wave function, and the formation of a finite conduction band below this threshold is not expected.

Quantitatively, one can calculate the localization length using the formula (6) in which the renormalized density of states calculated from the dispersion (11) is used. This approach is perturbative. The hypothesis made is that the inter-particle interaction is the main process leading to a well defined renormalized dispersion for the elementary excitations of the condensate, which is stable against the parametric processes. Second, the disorder is also a perturbation, which can only provoke scattering processes within this dispersion. In fact, there is a wide range of parameters [61], were both mechanisms should be taken into account simultaneously, leading to the instability of the condensate, something which we neglect here.

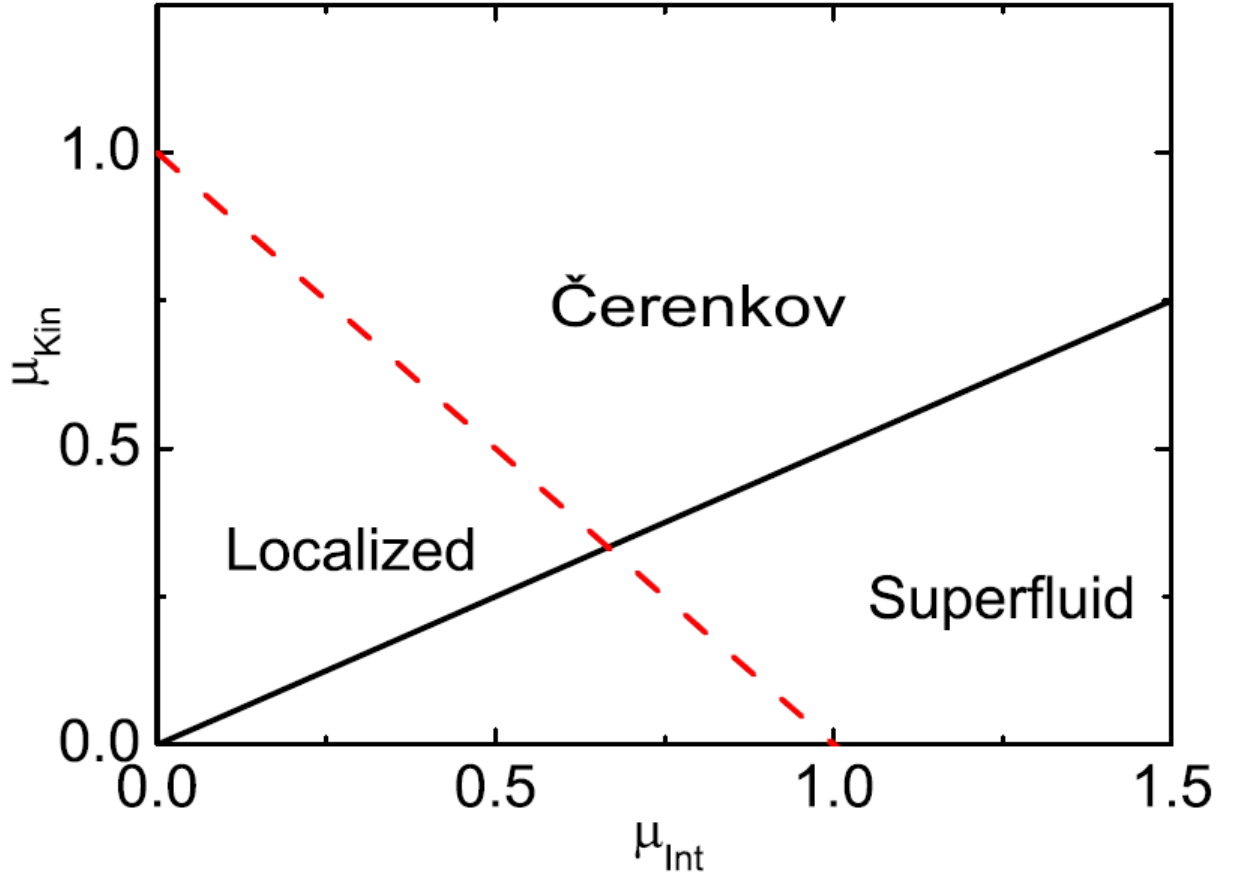

Figure 8: Phase diagram of condensed polaritons in the presence of disorder. $V_0 = 1$ meV, $m = 5 \cdot 10^{-5}\ m_0$. Everything below the red line is localised. Above the red and blue: Delocalised or Čerenkov. Above the red and below the blue: Superfluid.

For a parabolic dispersion (a valid approximation for polaritons with relatively small wavevectors), one can obtain a simple analytical expression for the density of excited states at the energy of the propagating condensate:

$$D_{1D}(E = E(k_0)) = \frac{1}{\frac{\hbar^2 k_0}{m} - \frac{\mu}{k_0}} \quad \text{if} \quad \left(\frac{\hbar^2 k_0^2}{m}\right) > \mu \qquad (14)$$

$$D_{1D}(E = E(k_0)) = 0 \quad \text{if} \quad \left(\frac{\hbar^2 k_0^2}{m}\right) < \mu \qquad (15)$$

Since the density of available states for the subsonic case is zero, $\gamma_{1D}$ drops to zero as well, and the condensate can propagate infinitely far without any scattering. For supersonic waves $\gamma_{1D}$ is given by:

$$\gamma_{1D}(k_0, \mu) = |x|^2 \frac{\sqrt{2} m^2 L_c V_0^2}{\hbar^4 k_0^2 - m\hbar^2 \mu} \Theta(\hbar^4 k_0^2 - m\hbar^2 \mu) \qquad (16)$$

where $\Theta$ is the Heaviside function

The denominator describes the enhancement of the scattering by disorder in the Čerenkov regime. One should note that $\gamma_{1D}$ goes to infinity close to the transition to a superfluid state which corresponds to a fully localized system. The hypothesis we made to derive the formula is that the disorder is acting as a weak perturbation on the propagating wave. This hypothesis breaks down close to the transition point between the superfluid and Čerenkov regimes.

In a system with a finite lifetime, a supplementary decay source should be considered and the total decay rate can be written as:

$$\gamma_{1D}(k_0, \mu) = |x|^2 \frac{\sqrt{2} m^2 L_c V_0^2}{\hbar^4 k_0^2 - m\hbar^2 \mu} + \left(1 - |x|^2\right) \frac{m}{\hbar k_0 \tau_R} \quad (17)$$

Figure 9 shows a plot of $\gamma_{1D}^{-1}$ which is the propagation length for interaction energy of 1 meV. In the low energy range, the kinetic energy is at least twice smaller than the interaction energy and polaritons are superfluid. Their propagation is only limited by their lifetime. When the speed of sound is reached (kinetic energy is one half of the interaction energy), the density of states available for elastic scattering processes passes from 0 to infinity and the condensate becomes very strongly localized. First, we can expect that this abrupt transition can be smoothed in a real system because, of finite linewidth, finite size effects and so on. Second, it is not clear at all if our approximation (stability of the condensate against parametric processes) remains valid in this case. The dashed line shows a plot of the decay in the non-interacting case (same as for figure 5) for comparison. The extra localization induced by the renormalization of the dispersion in the Čerenkov regime is clearly evidenced. It is interesting to notice, that for infinite life time particles our approach gives at the transition point a localization length passing from infinity to zero.

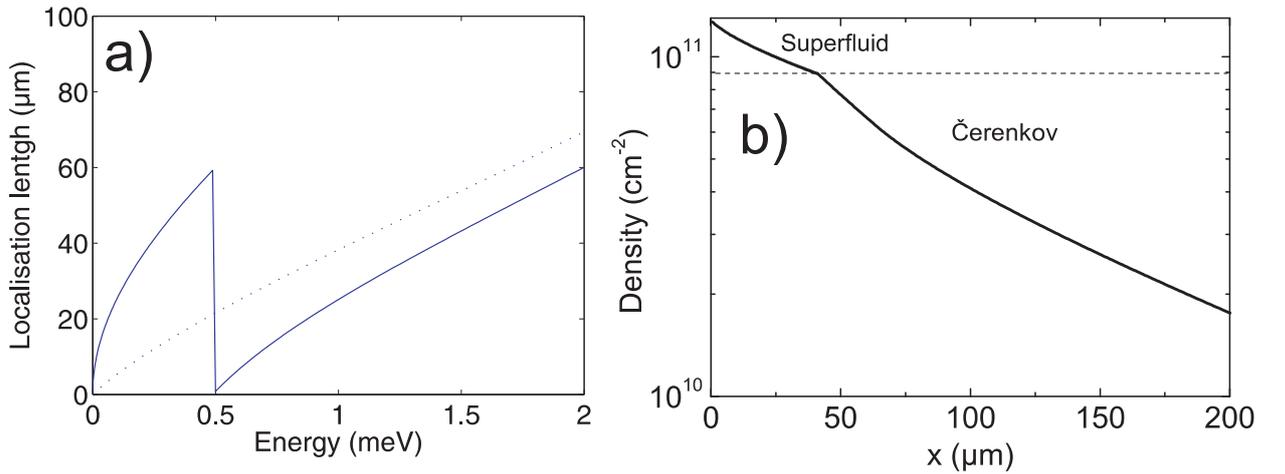

Figure 9. a) Propagation length of a 1D polariton condensate as a function of its energy in the presence of disorder $L_c = 1\mu m$, $V_0 = 0.25$ meV. Solid line: $\alpha|\psi_0|^2 = 1$ meV, dashed line, no interaction (same case as for figure 5)

b) Density profile of a 1D polariton condensate propagating in the presence of disorder. The transition from superfluid to Čerenkov regime is visible as a cusp.

We show the results of a hydrodynamic simulation of the propagation of a polariton condensate with a wavevector in a disordered system taking into account the finite particle lifetime and the backscattering induced by the disorder, both included in the equation (17). The condensate is injected (presumably by non-resonant pumping) at the point x=0 μm and propagates to the right. The potential *U(x)* felt by the condensate at the point *x* is composed of the disorder potential and the interaction energy linked with the local density. Figure 9b shows the steady-state situation, obtained when the interaction energy significantly exceeds the kinetic one at the injection point, and the condensate is therefore initially superfluid. Since the density is decreasing with the coordinate even in the superfluid regime because of the finite lifetime, the condensate propagates down a potential slope (created by the gradient of the density) and is therefore accelerated, until it reaches the (local) speed of sound. At this point, the condensate becomes supersonic and backscattering starts to play a major role, strongly reducing the localization length. When the potential energy is completely transferred into kinetic energy, the condensate starts to move faster and therefore the backscattering is reduced again. The localization length increases almost back to the value induced by the finite lifetime.

The transition between the superfluid and supersonic phases is a horizon at which strong instabilities may develop, but their study will be a subject of a separate work.

**V Conclusion and perspectives.**

The main result of this chapter is to calculate the localization length of an interacting polariton condensate, moving in a weak random potential. To our knowledge this type of calculation was never published before for this type of system. We consider the transition from a superfluid to an Anderson localized condensate. The localization length is found to be reduced in the Čerenkov regime. Till 2009, polariton lifetime was typically below 10 ps, and the propagation length of particles was typically limited by this factor. Another case realized is the one of large disorder were the condensate is classically localized. The fabrication of new structures, of high quality, and showing polariton life time of 30 ps [28] and even more [62] should allow to study in details localization and superfluidity phenomena. Another interesting direction relates to the realization of periodically modulated samples [63, 64]. In such type of structures, a wide variety of new phenomena (such as Bloch oscillations [65] for instance) can take place.